# THE ICT4D-BUEN VIVIR PARADOX: USING DIGITAL TOOLS TO DEFEND INDIGENOUS CULTURES

Lorena Pérez-García. Cross Cultural Bridges, perezgarcia@crossculturalbridges.org

**Abstract:** Arguably shaped by political economy perspectives from the Global North, ICT4D aims to reduce socioeconomic disparities across countries and regions through ICT implementations, as well as to open up opportunities for empowerment and human development. Despite these aims, ICT4D has been criticized because 1) although ICT and internet have positive effects on societies across the Global North, their positive impact on people's lives in the Global South cannot be easily proved; 2) ICT4D's primary focus seems to be on ICT's series of artefacts rather than on ICT's positive transformative potential of living conditions in the world; 3) the type of development ICT4D aims for could mask global hegemonic interests and seek neoliberal restructuring within less socioeconomically favoured communities within the Global South. For these reasons, claim scholars, ICT4D should be revised. By presenting ICT appropriations among Wixárika peoples in Mexico to protect their sacred land, this paper aims to 1) shed a light on the need for postcolonial critical frameworks on what 'development' associated with ICT should be and 2) to foster discussions on whether ICT can enable alternative voices from the Global South to be heard, despite tensions between traditional views and contemporary technologies.

## 1. INTRODUCTION

Technological advances achieved in microelectronics and telecommunications since the early 1990s resulted in the proliferation of information and communication technology (ICT). Furthermore, with the migration from Internet 1.0 to Internet 2.0, ICT became ubiquitous, as technology allowed more user involvement. While technological advances brought about technology-driven political changes and socioeconomic development, in some societies 'development' programmes resulted in a transfer of wealth away from low-income regions (Escobar, 1995). Thus, while ICT has become central to the global development agenda of economic growth and democracy (Wamala-Larsson et al., 2015), ICT has also contributed to socioeconomic and political inequality worldwide (Heeks, 2010) due to ICT's uneven distribution in the world.

To address this paradox, ICT for Development (ICT4D) emerged as an academic discipline and an institutional policymaking rhetoric to address inequality in developing regions and to enable economic development in developing and underdeveloped countries through ICT (Zheng et al., 2018; Wamala-Larsson & Stark, 2019). However, opting for the use of ICT to improve people's living conditions involves a paradox, according to Kleine (2010), as ICT and internet are highly regarded inventions that have had positive effects across the Global North, they are perceived as an adequate solution to many social problems. Yet when it comes to obtaining funding for ICT4D projects and programmes in the Global South mainly, it appears to be not that simple to prove their potentially positive impact on people's lives. Thus, institutionally speaking, the impact that ICT has on people remains an elusive issue (ibid). Moreover, the association of 'development' and ICT within the rhetoric of ICT4D has been criticised for masking global hegemonic interests and seeking neoliberal restructuring within less socioeconomically favoured communities (Svensson & Wamala-Larsson, 2015).

## 2. RESEARCH PROBLEM, CONTEXT AND METHODOLOGY

While there are institutional and governmental ICT4D projects successfully implemented worldwide, there are also institutional projects that fail to accomplish their goals (Pérez-García, 2020). Among the causes of failures of ICT4D projects, scholars stress that the type of development expected from ICT implementations seems to have a close relationship with economic improvement, and might be therefore rather restricted, as – although hardly imprinted in ICT4D projects and practices – development can be the result of the accomplishment of several material, personal and community expectations (Kleine, 2010; Sarjo et al., 2014; Tshivase et al., 2016).

When focusing in Mexico, although a few institutional ICT4D projects have succeeded in urban and reasonably urbanized areas, records show that most government ICT4D projects fail in rural areas. Among the accounted reasons for failure are 1) the economic and digital divide among the population of the country; 2) limited ICT access and ICT availability; 3) the difficulties to get ICT services in rural areas, due to geographic and natural conditions (i.e. mountains and biomass density). Linked to the effects these situations can have on ICT users' involvement, ethnic identity, literacy rates, digital literacy, and language ought to be counted as failure factors of ICT4D projects as well (Pérez-García, 2020).

In relation to ethnicity, scholars suggest that models of technological implementations may smuggle in neo-liberal values and goals of private wealth accumulation (Jiménez & Roberts, 2019). Also, growth-oriented neo-liberal development policies and ecological planning can be in some cases neo-colonial as well as environmentally unsustainable (ibid). Thus, uncritically applying approaches from the Global North in the Global South may subordinate indigenous knowledge, different ways of knowing, and diverse cultural values. Moreover, growth-orientated neo-liberal development could oppose to indigenous values, such as shared enterprise, communal interests, reciprocity and interconnectedness– all central to indigenous worldviews in Latin America (Kothari et al., 2016).

Regarding indigenous population in Mexico, over 50 different indigenous groups correspond to about 12% of the total population of the country. Pre COVID-19 pandemic, more than 70% of indigenous people lived in poverty conditions (about 40% in poverty, 32% in extreme poverty) (CONEVAL, 2016). Despite the structural violence associated to the poverty rate in which indigenous live, indigenous peoples use ICT – when available, and to some extent. By doing so they cover a basic need to communicate. Moreover, indigenous populations in Mexico, like any indigenous peoples from around the world, collaborate across social media and attract international support for their digital activism campaigns (Myers et al., 2020). Such campaigns focus mostly on restoring cultural identity and preserving natural resources (Young, 2018). Also, some indigenous groups may use information systems in ways similar to dominant society groups (e.g., to run tribal affairs related to finance, accounting, and member engagement). Yet, we should also consider that Indigenous groups and Westerners may conceptualize and use information systems in vastly different ways, which the academic literature does not always account for (Myers et al., 2020).

Thus –within a broader pluricultural effort to generate knowledge, to further understand ICT as a means to foster culturally diverse views on development and to bridge the digital divide in socioeconomically marginalised communities like the above described – this paper sustains the need to rethink ICT4D from a postcolonial perspective. By doing so, this research aims to contribute to document praxis and epistemological gaps between institutionalized ICT4D and successful indigenous ICT's use by replacing the expectations set on ICT for economic development with an indigenous post-colonial biocentric view.

To examine how ICT enables alternative voices and critical frameworks from the Global South to be heard, this paper will focus on ICT4D-Buen Vivir (BV) – good living– paradox that resulted from ICT appropriation among the Wixárika Indigenous community in Mexico. *BV* is a broad philosophical framework that stems from indigenous Quechua and Aymara traditions; being part of native Latin American populations' traditions, most contemporaneous indigenous views find support in it (Hernández-Díaz, 2018). Away from shamanism, animism and anthropocentric views on development, BV is a heterodox and hybrid view – already legally introduced in some Latin American countries – that comprises both traditional indigenous views and a critique to modernity. BV is an alternative order based on the coexistence of human beings across the spectrum of diversity and in harmony with nature (Hernández & Laats, 2020). It calls for a deep change of knowledge, affectivity (moral-social values) and spirituality (ethical values) to understand the relationship between humans and non-humans, without fostering separation between society and nature (Chuji et al., 2019). Altogether, these ideas could represent further possibilities for a substantial change *vis a vis* the current crises attributed to Western views on development, growth, maximization of production, competition, disproportioned use of natural resources, eco-destructive activities, minimization of the common good, and anthropocentrism. (cf. Hopwood et al., 2005; Yashar, 2005; Walsh, 2010; Vanhulst, 2015; De Sousa, 2014; Hollender, 2014, Gudynas, 2014; Chuji et al. 2019).

Although there is no singular view of BV among indigenous cultures in Latin America, traditional expressions referring to a 'good living' can be found throughout indigenous groups in the continent. In Mexico, indigenous groups appeal for *lekilaltik*, the 'good of us', from the Tojolabal people; *lekil kuxlejal*, 'the good life', from Tseltal populations; or Maseual's *yeknemilis,* 'good life' (Hernández-Díaz, 2018). Although in this paper a case study from Mexico will be presented, Buen Vivir will be considered and no Mexican expression that appeals for the same as – in its general form – BV has already been linked to the creation of legal frameworks for the rights of nature, such as Ecuador's political constitution, the European debate on happiness, well-being, the critique of economic growth, and has also been used in economic development programmes in Mexico. As these principles could provide alternatives to current development and environmental crises (cf. Pérez-García, 2020; Hernández & Laats, 2020), finding the link between ICT4D and Buen Vivir could hold considerable elements for reflection when aiming to examine how ICT enables alternative voices and critical frameworks from the Global South to be heard. To accomplish this goal, a qualitative interdisciplinary research with plural methodology with a post-development paradigm has been carried out. As a result, this paper is divided into four sections, as follows: Firstly, the literature to date on development associated with ICT4D will be reviewed, looking specifically at how neoliberalism, globalisation, development, and ICT are being represented as a plausible solution socio-economic inequality. Although not yet broadly associated with ICT4D, Buen Vivir will be taken into consideration. Moreover, the indigenous pursuit of postcolonial and culturally inclusive notions of development through ICT will be introduced with ethnographic research carried out to document the use of ICT by Mexican Wixárika indigenous people. Drawing upon these sections, the findings of the paper will be then presented into two parts, particularising first the lessons learned from the presented case and, secondly, elaborating further the need to use ICT to enable alternative voices from the Global South to contribute with notions of development. Finally, overall conclusions will be presented.

## 3. DEVELOPMENT AND SUB-ORDINATION
Development as an amelioration has been present for centuries as societies strived for quality of life. As societies became more complex overt time, the triggers and range of development enlarged, rendering development into multidimensional ethos and praxis. However, in contexts where Western culture dominates, a form of colonisation takes place through social constructions of meaning around a rigid hierarchy in which the colonisers outrank the colonised (Ashcroft, Griffiths, & Tiffin, 1998).

Development's origins as progress, rationalism and modernity are rooted to European thinking (Rist, 2002; Cowen & Shenton, 1996; Hopper, 2013). More contemporary views on development were forged in aftermaths of the World Wars, as victorious nations aimed for post-war reconstruction and economic stability (Hopper, 2013), bringing about change within societal structures, people and culture as well, expecting to replicate the values of the First World into the Global South (Sachs, 1992; Sheppard et al., 2009)

By the Cold War era, the aim for development reached technology, as another agent against communism. Governmental/military projects were developed –i.e. the ARPANET (Advanced Research Projects Agency Network) and satellite communications networks in collaboration with the UK and France – involving investment from several countries and expectations for diverse geographies, mainly within the First World (Naughton, 2016).

By the mid-70s, after both the oil crisis and the economic crisis broke, the light was shed on the fact that the economic models promoted during post-war development were unsustainable in the long run for not taking any account of the effect that production and oil extraction could have on the environment and, consequently on the society (Sheppard et al., 2009). Thus, as industrialization started to reach its limits within first world countries, new markets were sought.

Neoliberalism appeared as an alternative. With neo-liberal economy free trade in commodities, services, and the international movement of capital and labour, politicians and economists (mainly Anglo-Americans) expected that every person in the world, regardless of their socio-economic background, would cultivate the same intention for 'development' and opportunities to prosper. Development was thus understood within the Global North as a market-led personal prosperity plan (Sheppard et al., 2009). An appeal for globalisation came then into play. Under this scheme, individuals were expected to experience the maximal personal liberty that resulted from market possibilities and deregulation of the private sector. Freedom was thus understood as a benefit of market-led neo-liberal economy, as individuals would experience it, as long as the market and the economy of the area they lived in would allow it (ibid). With such powers endowed to globalisation, developed countries leaned towards believing that globalisation and neoliberalism would be able to solve the economic difficulties in developing and underdeveloped countries. In consequence, political and economic pro-globalisation forces aimed for establishing the neo-liberal globalisation model in the largest part of the world.

In spite of this, the consequences of market-led developed economies started to take a toll on developing and underdeveloped societies. Alternatives for market processes to alleviate poverty, underdevelopment, and to counter the subordination brought about by globalisation and its negative effects in developing and underdeveloped societies, sustainable development emerged (Sheppard et al., 2009). Within this framework, nature preservation, on the one hand, and social capital, on the other hand, started to be acknowledged.

While this was happening, ICT use and applications were displaying an accelerated growth in some parts of the world. To bridge the socioeconomic gap, global institutions aimed to address inequality issues in the world with ICT towards the arrival of the new millennium (Unwin, 2009). Efforts between the Organisation for Economic Co-operation and Development (OECD) and the United Nations (UN) eventually resulted in the Millennium Development Goals (MDGs), a set of eight goals to be accomplished by the target date of 2015.

Regarding sustainable development, in 2012, the UN invited stakeholders to make voluntary commitments on delivering concrete results for sustainable development. The Conference on Sustainable Development – Rio+20 – took place, drafting clear and practical measures for sustainable development. Ventures for a paradigm shift from a human-centric society to an Earth-centred global ecosystem – derived from this conference – are imprinted in current socio-economic and political efforts from countries like Iceland, Scotland, New Zealand, France, Netherlands, and Wales, to achieve economic degrowth for ecological sustainability and social equity through:
1. alternatives to gross domestic product (GDP) as a measure of well-being,
2. radical questioning and proposals on common values of care, solidarity, cooperation and an understanding of Nature,
3. collaboration of national and regional governments to advance on the recognition of development as both human and ecological well-being (UN, 2020).

By 2014, the OECD and the UN realised that the MDGs were far from being attained. To address these shortcomings, the 2030 Agenda for Sustainable Development was agreed in September 2015, replacing the MDGs by the Sustainable Development Goals (SDGs). ICT was identified as a loyal ally "to ensure that 'no one is left behind' in the course of progress towards sustainable (economic) development" (ITU 2016, pp. 79). Despite this re-evaluation, still nowadays over 40% of the global population has no access to internet and ICT (Kemp, 2021), close to 46% of the world's population lives on less than $5.50 USD a day, and close to 26% lives on less than $3.2 USD a day (WB, 2018).

This prevalent digital and social divide highlights that despite institutional efforts, ICT could not shift social structures of power and would not translate on social difference (De Sousa, 2006; Appadurai 2013). Thus, critical voices pointed out the proliferation of epistemological gaps. In this regard, claimed De Sousa (2006), globalisation and its associated practices disqualify people – artificially and epistemologically – to maintain the domination of neoliberalism, globalisation and their techno-scientific grounds based upon people's alignment to alternative types knowledge – such as traditional indigenous knowledge. People with alternative knowledge are rendered "invisible, unintelligible or irreversibly discardable", resulting in a creation of non-existent subjects and objects (De Sousa, 2006, p.16), being ethnically diverse populations among the most affected.

The socioeconomic and political discrimination rooted to globalisation-imposed subalternity and non-existence would keep ethnic diversity from being considered as starting point on the elaboration of comprehensive views of 'development' and as a trigger of ICT implementations to boost national economies (Navarrete 2016, 2017). While non-existence takes place, rough poverty, marginalization, their effects and consequences, would lead people to develop a culture of poverty (Appadurai, 2013). This 'culture' would act as a constant reminder among the poor of what they cannot attain and would aim for the acknowledgment of poverty to gain back a little of the dignity lost.

Under this rationale, indigenous have been rendered as subaltern and, as such, their culture and their cultural artefacts would also be considered non-existent within a globalisation paradigm. Given that most indigenous peoples are vulnerable due to the colonial oppression's lingering effects, institutional ICT implementations relates to the ethics and politics of engagement in research, especially as colonialism has marginalized many indigenous people to the extent that still nowadays they continue to face threats to their sovereignty, wellbeing, and natural environment (Myers et al.,2020). Despite this situation, ethnic groups are not frozen in time, living a life without leisure or fun, and oppressed; they are as much members of the digital generation as persons living in the Global North (Gajjala, 2014). Thus, when available, ethnic populations are active ICT users, as shown by documented cases of ICT use among indigenous communities – triggered by themselves – in which through ICT appropriations, they have been able to pursue their own goals, show their resilience and stress the need to reassess ICT4D, to include postcolonial views on development and to favour the environment over economic gain (Pérez-García, 2020) and to reach an understanding of ethical ways to include indigenous peoples in ICT research process and consequent benefits without dictating the ways in which we include indigenous peoples (Myers et al.,2020).

# 4. THE WIXÁRIKA USE OF ICT TO SAVE SACRED LANDS

To approach indigenous ICT initiatives, qualitative research was carried out, focusing on the Wixárika (Huichol) ICT use and appropriation as a means to protect their sacred land from development projects (mining, agriculture, tourism) and their call for a fairer behaviour towards the environment. Between 2011 and 2019 diverse digital ethnography methods were carried out[1] to document the activities carried out by the Wixárika people in Mexico (associations and individuals) to stop mining and touristic projects in their sacred lands. With the information gathered, the evolution and progress of indigenous and grassroots initiatives through their online presence was documented. Moreover, with consent of social media users (Facebook, Twitter, YouTube), the activities of 5 indigenous associations on their accounts and their websites were periodically monitored. Also 30 structured and semi-structured interviews to 15 NGO workers and indigenous peoples were carried out online, in Mexico and in the Netherlands. Some of the interviewees where interviewed more than once over time, without exceeding three times. Note that as the focus of this paper is on the movement to save Wirikuta, the information presented and analysed corresponds to the period between 2012-2016. Nevertheless, although not approached in this paper, the research on Wixárikas' use of ICT continues to further document the ramifications of the movement, further achievements and their influence on other indigenous populations in Mexico.

Wixárika people – Huichols in Spanish or *Wixaritari* in Huichol language – hold nowadays the closest traditions to those of the Aztecs, hence their cultural importance (Liffman, 2011). One of the many particularities of this indigenous group is that, despite being active participants in the socioeconomic and politic life of the country, they have preserved their spiritual identity, continuing with their practices of animism – a cultural and religious tradition going back over 2,000 years (Alarcón-Cháires et al., 2013).

Wixárikas corresponding to about 4% of the total indigenous populations of the country (INEGI, 2010). Half of them is spread throughout their traditional lands in Nayarit, San Luis Potosi, Jalisco, Zacatecas and Durango. The other half is scattered throughout the rest of the country or in the USA (CHAC, 2011). The language spoken by them is *Wixaritari,* and several dialects derived from it. While the majority of them speak Spanish, some members remain mono-linguistic. Although Wixaritari is not a written language, Wixárikas are convinced that their traditional myths and traditions can only live on orally thus the passing on of the language is essential for the community's survival and the development of the Wixaritari identity (Fernández, 2013).

One of their cultural pillars is a close spiritual relationship with their ancestors and their deities. To communicate with them, the Wixárikas make at least five pilgrimages through their sacred places – Wirikuta being the most important. By pilgriming, Wixárikas adhere to their beliefs on the creation of the world and connect with their deities to receive protection and guidance through life (Liffman, 2011).

In the Wixárika cosmovision, Wirikuta it is the place where the world was created, as it is one of the ecosystems with the highest number of endemic species in the world (ibid). Thus, from childhood every Wixárika is expected to make a pilgrimage to Wirikuta to re-enact the path their spiritual ancestors followed to communicate with their deities. Not performing these pilgrimages and not being in contact with nature in their sacred places "would separate the Wixaritari from their divine beings, resulting in a life without purpose" as in Wirikuta "the essence of terrenal [worldly] life is weaved and sustained" (Alarcón-Cháires et al., 2013).

---

[1] Such as digital participant observation, online interviews, interaction observation on social media platforms and normalised web distance among related terms (see Pérez-García et al., 2016).

Despite the complex particularities of Wirikuta, as the 2008-2009 global economic crisis hit Mexico, the federal government decided to boost the country's economy with mining projects in the land, granting in 2009 78 open-cast mining concessions to Canadian companies in the state of San Luis Potosí, affecting at least 70% of Wirikuta (ibid).

In response, Wixaritari authorities formed a Council for the Defence of Wirikuta, RWCDW in 2010. The Council, with the support of national and international NGOs, founded a Front in Defence of Wirikuta, FEDDW. The Front tried through different avenues to reach representatives of the mining companies in Canada and Mexican federal government officials. With little response and achievements, they hoped to get support from the UN, as the mining concessions were about to start their operations. The Front changed its strategy and focused instead on creating digital content to be available on social media so that the online community could "visualise and broadcast" — nationally and internationally — the needs of the Wixárikas and show solidarity for the defence of Wirikuta (Alarcón-Cháires et al., 2013). Throughout the online content made available – mostly recorded in Wirikuta –, Wixárikas presented the viewers and readers with information about their culture, their land, their pilgrimages and the way in which the mining concessions threatened their cultural heritage and the ecosystem. Information was recorded in Wixaritari by the indigenous people and translated into Spanish by the Frente. Soon, translations were also made available in English, as well as in French, German and Italian.

The spread of information via social media, the support from international personalities in the arts, literature, activism, and the response from national and international organisations led – in February 2012 – to the suspension of one of the mining projects that would have carved up 3% of Wirikuta (DOF, 20/02/2012). Three months later, in May, the federal authorities announced that Wirikuta was to be spared from the mining and returned to the indigenous communities (Presidencia de la República, 24/05/2012). Huichol indigenous authorities argued that half of the territory was still going to be exploited by the mining companies (Venado Mestizo, 25/05/2012).

Further pressure was exerted through social media releases and the further involvement of indigenous ICT users. In September 2013, the pressure applied by the online movement to Save Wirikuta was enough to force the federal government to suspend – not cancel for good – all the mining concessions in Wirikuta (DOF, 13/09/2013). With this resolution at least 500 direct jobs and 1,500 indirect jobs were cancelled, along with a planned investment of over 17 million MXP (approx. 1 million euros) (LGDP, 2014). Wirikuta was, nevertheless, temporarily spared from the ecological damage that mining would bring about and became a biosphere reserve for the time being, protecting not only the Wixárika traditions but the endemic species of the area.

## 5. FINDINGS

### 5.1 ICT and indigenous people

Despite the socioeconomic conditions of indigenous people in Mexico and the scarce ICT possibilities in the areas where the Wixárika initiative developed, the involvement of the population, the wide use of ICT to spread content, and the high impact of the initiative outweighed the odds against them. Through interviews to Wixárika Facebook users, light was shed on the fact that for the content to be uploaded and transferred, users committed to travel to mountains' tops, up to areas where it is already known that there is better mobile signal reception. Although there were NGOs backing up the indigenous people with their initiative, indigenous people where actually generating digital content for websites and various social media accounts to achieve group's recognition, cultural preservation, and to protect the ecosystem where they live.

The most valuable lessons learned from the presented case are:

1. While fulfilling a basic need to communicate with ICT, the Wixárika's ICT appropriations contributed to the group's quest for recognition, autonomy, ecological balance, cultural preservation and survival of their ethnic origins.
2. Despite the barriers and issues that Wixárikas encountered to use ICT, their ICT appropriation presented them with a means to achieve community goals that – while legally and ethically defendable – do not correspond to national governments and national and international institutions' planning.
3. When populations living under structural violence interact with ICT –like Wixarikas in the presented case–, it is important to draw upon the experiences of these populations to understand what is – for them – to be pursued with ICT, whether appropriations and affordances could lead to strong ICT planning and implementations, and whether the latter could be a means for the fulfilment of their expectations on the relationships in their community and the relationship they develop with the environment.
4. Indigenous peoples seek to add their voices and aspirations to the debates on their future and their expected 'development' through ICT implementations in order to change how they are perceived by the world, to preserve their culture, to improve their living conditions while living within tradition.
5. The appropriations and affordances Wixárika population made of ICT exemplify how ICT can be used to achieve nature conscious improvements in living conditions, without relying on economic growth only.

## 5.2 ICT and bio-centric views on 'development' from the Global South

Most societies in the world live within a globalisation paradigm. However, the wealth and well-being that deregulated free trade would bring about for everyone involved, has not been homogenously distributed. Moreover, Western domination through globalisation has profoundly marginalised knowledge and wisdom that had been in existence in the Global South (De Sousa, 2014). Thus, to the eyes of populations of the Global South, development as seen in the Global North is a rigid idea that stacks nations – and people – into categories. It separates nature from humans and humans in gender, religion, class (Bassey, 2019). In contrast, indigenous traditions, claims Gudynas (2015), do not involve a linear idea of progress and history. Moreover, indigenous ontologies see the world as a plurality of simultaneous stories, with no single totalising narrative, where humankind is closely interrelated with nature (Jiménez & Roberts, 2019; Hernández & Laats 2020).

It would be overly ambitious to generally assert that ICT could help to change the ways the less dominant sociocultural groups are globally understood. Nevertheless, the paradox depicted in the presented Wixárika case, – along with other cases, such as Aymara Women's use of ICT to preserve traditional knotting patterns (Jiménez-Quispe, 2021), Pastoralists use of ICT to lead their herds through paths with food and water (Bodgan-Martin, 2021) – shed light on the fact that ICT indigenous people's appropriations, affordances, and use – far from only achieving economic ameliorations – are a means to expand people's capacity to exercise their voice, to debate the economic conditions to which they are confined, to protect their culture, and to redefine the way they want to live their lives. Understanding the agents, actors, and tensions involved in this type of cases could help us to understand that the current development paradigm associated to ICT4D could be excluding not only segments of society but also efforts, – even if these are not fully valued or understood in the Global North, like the presented case – that aim for an improvement on people's living conditions and preservation of the environment. Also, by understanding the ICT4D-Buen Vivir paradox and whether different views on 'development' as an amelioration could be associated with ICT4D could represent change possibilities to stir Western anthropocentrism views on development as production, use of natural resources and minimization of the common good, supported by ICT4D practices, and the effect these have had in contemporary societies. Moreover, it could minimise the effects of the logics of non-existence and the culture of poverty described by De Sousa (2006) and Appadurai (2013) as BV rejects all forms of colonialism upholding a type of interculturality that values each tradition of knowledge, incorporating reciprocity, complementarity, communalism, redistribution (Chuji et al., 2019).

# 6. CONCLUSIONS

Reality, claim Jiménez & Roberts (2019), is constituted by many kinds of worlds, many ontologies, many ways of being in the world, many ways of knowing reality, and experimenting those many worlds. However, since colonial times, certain kinds of knowledges have been privileged as valid, consequently prioritising certain activities, while delegitimising and subordinating others with less economic and sociopolitical power (Tuhiwai-Smith, 1999; De Sousa, 2006, 2014).

When referring to ICT research, many new theories, claim Myers et al. (2020), are still grounded in colonial views. As a result, the research could lack sensitivity to indigenous peoples' issues, and enforce a materialism that does not correspond necessarily with indigenous ethos.

To advance knowledge on indigenous ICT uses research, argue Tsibolane in Myers et al. (2020), researchers should engage with indigenous communities to re-centre indigenous voices, languages, concepts, worldviews, histories, experiences, knowledge, and beliefs to address ICT paradigms while also referring to the ethical interdependence between human beings, the natural environment, and ancestors.

Regarding Buen Vivir, the experiences of Ecuador and Bolivia's nation-state constitutional projects demonstrated that BV can become co-opted as a discourse, without much change (Gudynas 2011; Williford 2018) and that there is a lack of structural preconditions for the implementation of BV at a nation level Beling et al. (2018). Other authors argue (Kothari et al., 2014; Merino,2015) that given that BV has been dialogically incorporated into government initiatives, its results and implications should be measured 1) by its contribution to destabilising dominant existing socio-cultural templates, and 2) by the possibilities it opens to explore alternatives to an economically neo-liberal way of living.

Although it can be claimed that BV is still a discursive 'work-in-progress', result from scholars and political leaders' interpretation and articulation of traditional indigenous knowledges done to present an alternative to development (Jiménez & Roberts, 2019), BV's strength resides in the fact that it does not aim to become a dominant hegemonic ideology. BV aims for the recognition of multiple perspectives coexisting and proposes to replace individualism values and economic growth at an environmental cost, with solidarity, reciprocity, complementarity, interdependence and harmony with nature. Considering such values could provide strong alternatives to perform ICT research that allow us to shift from a Eurocentric hegemonic ICT4D to a decolonial ICT4D that encourages plural ways to understand the world, while preserving nature.